\documentclass[12pt,a4paper]{article}
\usepackage{epsfig}
\begin{document}
\title{An embedding potential definition of channel functions}
\author{J. E. Inglesfield,\\
School of Physics and Astronomy,\\ Cardiff University, 
Cardiff, CF24 3YB, UK \\S. Crampin,\\Department of Physics,\\
University of Bath, Bath, BA2 7AY, UK\\ and H. Ishida,\\
College of Humanities and Sciences,\\ Nihon University, 
Sakura-josui, Tokyo 156-8550, Japan}
\date{}
\maketitle
\begin{abstract}
We show that the imaginary part of the embedding potential, a generalised
logarithmic derivative, defined over the 
interface between an electrical lead and some conductor, 
has orthogonal eigenfunctions which define conduction channels into and out of
the lead. 
In the case of an infinitely extended interface we establish the relationship 
between these eigenfunctions and the Bloch 
states evaluated over the interface. Using the new channel functions, 
a well-known result for the total transmission through the conductor 
system is simply derived.
\end{abstract}
\section{Introduction}
In scattering theory, and consequently in the theory of electrical
conductance \cite{butt1}, the concept of channels is 
fundamental -- channels are the asymptotic states at 
a particular energy, identified by some quantum numbers, 
between which the electrons are scattered. 
In conductance studies, which is our main concern, 
the current enters a nanostructure (for example) through open channels
in the leads on one side, and is transmitted into the open channels on
the other side \cite{butt2}. The theoretical problem in finding the 
conductance then becomes one of calculating the transmission 
probability between the channels \cite{fisher}. In this paper we show
that channels can be usefully defined in terms of the embedding
potentials which embed the nanostructure on to the leads \cite{jei1,jei2}. 
Moreover, this gives a simple derivation of the well-known result
\cite{levy,damle,wort} 
that the total transmission through the nanostructure between the
left- and right-hand leads at energy $E$ is given by
\begin{equation}
T_{lr}(E)=4\textrm{Tr}[G_{lr}(E)\Im\textrm{m}\Sigma_r(E)G^*_{rl}(E)
\Im\textrm{m}\Sigma_l(E)]. \label{trans1}
\end{equation}
Here $G_{lr}$ is the Green function connecting the left- and
right-hand surfaces of the nanostructure, and $\Sigma_{l/r}$ is a 
self-energy, or embedding potential, which couples each surface to the
left- or right-hand contact respectively. The trace is over the
quantum numbers of the channels in each contact. It was Wortmann 
\emph{et al.} \cite{wort} who realised that
$\Im\textrm{m}\Sigma_{l/r}$ in (\ref{trans1}) is the same as Inglesfield's
embedding potential \cite{jei1}, and derived (\ref{trans1}) in the
embedding context. They have used this result with embedding
to study the transmission of metallic interfaces \cite{wort}, 
and recently field emission has been calculated using the same 
formalism \cite{hiroshi}. 

The embedding potential in (\ref{trans1}) has been widely used in 
surface, interface, and transport calculations 
\cite{jei2,ishida,dix}. In surface
calculations, for example, the surface region is joined on to a
semi-infinite substrate, which can be replaced by an embedding
potential. The embedding potential, added on to the Hamiltonian for
the surface region, ensures that the surface wave-functions match onto
the substrate wave-functions. This means that the whole problem can be
treated by solving the Schr\"odinger equation for one or two layers of
atoms at the surface rather than the whole semi-infinite substrate.

In this paper we shall show how the embedding potential over a surface
can be used to define channels crossing the surface. We shall relate
these channels for the case of an infinitely extended two-dimensional
surface of a semi-infinite solid to the Bloch states in the solid. Our
results enable us to derive some interesting formulae in scattering
theory, and in particular allow us to derive (\ref{trans1}) rather
easily.

We use atomic units (a.u.), in which $e^2=\hbar=m_{\textrm\scriptsize e}
=1$, and also eV for energy (27.2 eV = 1 a.u.).
\section{The embedding potential and flux}
The embedding potential $\Sigma$, which is the basis of this paper, is a 
generalised logarithmic derivative for the substrate region which 
it replaces \cite{jei1}, satisfying
\begin{equation}
\frac{\partial\psi}{\partial n_S}(\mathbf{r}_S)=-2\int_S d^2r'_S
\Sigma(\mathbf{r}_S,\mathbf{r}'_S)\psi(\mathbf{r}'_S). \label{embpot1}
\end{equation}    
Here the integral is over the boundary $S$ between the surface and
substrate, and in the partial derivative $n_S$ is the normal to $S$,
taken into the substrate; $\psi(\mathbf{r})$ is the solution of the 
Schr\"odinger equation in the substrate, integrated from some 
boundary value $\psi(\mathbf{r}_S)$ on $S$ with outgoing boundary
conditions into the substrate. $\Sigma$ is the same as the 
mathematicians' Dirichlet to Neumann map \cite{isakov}.

The embedding potential can be found from $G_0$, the substrate 
Green function satisfying a zero-derivative boundary condition on $S$
\cite{jei1},
\begin{equation}
\Sigma(\mathbf{r}_S,\mathbf{r}'_S)=G_0^{-1}(\mathbf{r}_S,\mathbf{r}'_S),
\label{embpot2}
\end{equation} 
where $G_0^{-1}$ is the inverse of $G_0$ over $S$. The surface inverse
can be avoided by writing $\Sigma$ in an alternative form, in terms 
of a double normal derivative of the Green function $\hat{G}$ 
satisfying the zero amplitude boundary condition \cite{afisher},
\begin{equation}
\Sigma(\mathbf{r}_S,\mathbf{r}'_S)=-\frac{1}{4}\frac{\partial
\hat{G}}{\partial n_S\partial n'_S}(\mathbf{r}_S,\mathbf{r}'_S).
\label{embpot3}
\end{equation}
In this paper we shall be considering the embedding potential 
as a surface operator in the real-space representation of the
Hamiltonian, but it can also be found in tight-binding
form \cite{jei2,owen}, where it is usually called a self-energy. 

The current can be written in terms of $\Im\textrm{m}\Sigma$, which 
is ultimately why this term appears in the transmission expression 
(\ref{trans1}). To show this, we start from the expression for the
current density
\begin{equation}
\mathbf{J}=\frac{1}{2i}\left\{\psi^*\nabla\psi-\psi\nabla\psi^*\right\}.
\label{current1}
\end{equation}
We now consider an embedding surface $S$, the interface, say, between 
the nanostructure and one of the leads (figure 1). 
\begin{figure}
\begin{center}
\epsfig{width=12cm,file=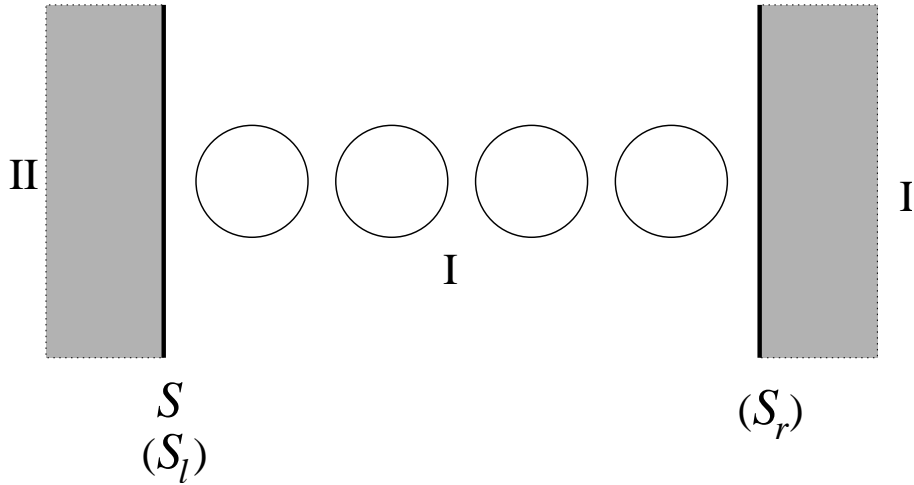}
\end{center}
\caption{Nanostructure joined on to metallic leads, represented by 
shaded areas. Region I, consisting of the nanostructure + right-hand 
lead is embedded over $S$ onto the left-hand lead, region
II. Subsequently it is convenient to label the left- and right-hand
interfaces by $S_l$ and $S_r$.}
\end{figure}
Substituting (\ref{embpot1}) into (\ref{current1}), the flux across
$S$ is given by
\begin{equation}
I=i\int_S d^2r_S\int_S d^2r'_S
\left[\psi^*(\mathbf{r}_S)\Sigma
(\mathbf{r}_S,\mathbf{r}_S')\psi(\mathbf{r}_S')-\psi(\mathbf{r}_S)
\Sigma^*(\mathbf{r}_S,\mathbf{r}_S')\psi^*(\mathbf{r}_S')\right].   
\label{flux1}
\end{equation}
Note that because of the convention for the sign of the normal 
derivative in (\ref{embpot1}), this is the flux from the 
nanostructure into the lead. As the embedding potential has the 
symmetry property of Green functions, namely
\begin{equation}
\Sigma(\mathbf{r}_S,\mathbf{r}_S')=\Sigma(\mathbf{r}_S',\mathbf{r}_S),
\label{embpot4}
\end{equation}  
equation (\ref{flux1}) simplifies to \cite{wort}
\begin{equation}
I=-2\int_S d^2r_S\int_S d^2r'_S \psi^*(\mathbf{r}_S)\Im\mathrm{m}
\Sigma(\mathbf{r}_S,\mathbf{r}_S')\psi(\mathbf{r}_S').\label{flux2}
\end{equation}
The fact that (\ref{flux2}) gives the flux \textit{into} the lead is
associated with our convention that $\Sigma$ corresponds to outgoing
waves. This is because $G_0$, whose surface inverse gives the
embedding potential (\ref{embpot2}), is evaluated at energy
$E+i\epsilon$, where $\epsilon$ is a positive infinitesimal.   

At this stage we introduce the eigenfunctions of $\Im\mathrm{m}\Sigma$
at each energy as a natural choice of channel functions,
given by the equation
\begin{equation}
\int_S d^2r'_S \Im\mathrm{m}\Sigma(\mathbf{r}_S,\mathbf{r}_S')
\psi_i(\mathbf{r}_S')=\lambda_i\psi_i(\mathbf{r}_S).\label{eigen}
\end{equation}
Although (\ref{eigen}) only gives the channel function 
$\psi_i(\mathbf{r}_S)$ over $S$, this uniquely defines the 
corresponding outgoing wave-function throughout the substrate. Because 
$\Im\mathrm{m}\Sigma(\mathbf{r}_S,\mathbf{r}_S')$ is real,
and symmetric in $\mathbf{r}_S$ and $\mathbf{r}_S'$, the eigenvalues
are all real. The eigenfunctions can be taken to be real and 
orthogonal. If the eigenfunctions are normalised to unity over $S$,
\begin{equation}
\int_S d^2r_S \psi_i(\mathbf{r}_S)^2=1, \label{norm}
\end{equation}
we see by substituting into (\ref{flux2}) that the flux associated
with $\psi_i$ is $-2\lambda_i$. This must be positive or zero, because
of the outgoing boundary conditions implicit in the embedding
potential, so we conclude that $\Im\mathrm{m}\Sigma$ is negative 
semi-definite, with eigenvalues negative or zero. The eigenfunctions 
with non-zero eigenvalue correspond to flux-carrying, open channels; 
there is an infinite number of closed channel eigenfunctions satisfying
(\ref{eigen}) with zero eigenvalue. 

$\Im\mathrm{m}\Sigma$ can be expanded in the usual way, as a sum over
its eigenfunctions, and with the normalisation given by (\ref{norm})
we have
\begin{equation}
\Im\mathrm{m}\Sigma(\mathbf{r}_S,\mathbf{r}_S')=\sum_i \lambda_i
\psi_i(\mathbf{r}_S)\psi_i(\mathbf{r}_S'). \label{eigsum1}
\end{equation} 
This reduces to a sum over the open channel functions,
\begin{equation}
\Im\mathrm{m}\Sigma(\mathbf{r}_S,\mathbf{r}_S')=
\sum_{\mbox{\scriptsize open}~i}  \lambda_i\psi_i(\mathbf{r}_S)
\psi_i(\mathbf{r}_S'), \label{eigsum2}
\end{equation}
a result which we shall find useful later on.

In conductance, it is only the open channels which are important, 
those with non-zero eigenvalue. The same open channel functions 
can be used for flux \textit{out} of the lead into the nanostructure; 
this corresponds to using the $\Sigma$ appropriate to incoming waves, 
which just involves a change in the sign of $\Im\mathrm{m}\Sigma$.   
These channels have the advantage of being defined via (\ref{eigen}) 
over the interface
between the lead and the nanostructure, precisely where they are needed.
They do in fact correspond to the usual channels in at least one case,
the case of waveguides. With a waveguide lead, the separability of the 
Schr\"odinger equation for motion along the waveguide and across 
the guide leads to different waveguide modes, and an embedding 
potential which is diagonal in a mode representation \cite{dix}. As a result, 
each embedding potential channel function $\psi_i(\mathbf{r}_S)$ 
corresponds to a waveguide mode.   

\section{Bloch states and channel functions}
In the case of an infinitely extended interface, it is interesting 
to study the relationship between our channel functions and the 
substrate Bloch states evaluated over this interface. With an 
extended interface, the ``lead'' on one side consists of a semi-infinite 
metal in which all wave-functions can be labelled  by the Bloch wave-vector 
parallel to the interface, $\mathbf{K}$. 
The embedding potential can be expanded as
\begin{equation}
\Sigma(\mathbf{r}_S,\mathbf{r}_S')=
\sum_\mathbf{K} \sum_{n,m} \Sigma_{\mathbf{K},nm}
\chi_{\mathbf{K},n}(\mathbf{r}_S)
\chi^\ast_{\mathbf{K},m}(\mathbf{r}_S')
\end{equation}
using a set of orthonormal surface functions 
$\chi_{\mathbf{K},n}(\mathbf{r}_S)$, in general complex, that satisfy 
Bloch's theorem with wave-vector $\mathbf{K}$. Using
(\ref{embpot4}) it follows that 
\begin{equation}
\Im\mathrm{m}\Sigma(\mathbf{r}_S,\mathbf{r}_S')=
\sum_\mathbf{K} \widetilde{\Sigma}_{\mathbf{K}}(\mathbf{r}_S,\mathbf{r}_S')
\end{equation}
with $\widetilde\Sigma_{\mathbf{K}}$ expanded as
\begin{equation}
\widetilde{\Sigma}_{\mathbf{K}}(\mathbf{r}_S,\mathbf{r}_S')=
\sum_{n,m} \widetilde{\Sigma}_{\mathbf{K},nm}
\chi_{\mathbf{K},n}(\mathbf{r}_S)
\chi^\ast_{\mathbf{K},m}(\mathbf{r}_S')
\end{equation}
where
\begin{equation}
\widetilde{\Sigma}_{\mathbf{K},nm}=\frac{1}{2i}\left[
\Sigma_{\mathbf{K},nm}-\Sigma^\ast_{\mathbf{K},mn}
\right].
\end{equation}
$\widetilde{\Sigma}_\mathbf{K}$ is Hermitian and so has real eigenvalues, 
and its 
eigenfunctions are the channel functions since the flux across the surface $S$
of an outgoing state with Bloch wave-vector $\mathbf{K}$
\begin{equation}
\psi(\mathbf{r}_S)=\sum_n c_{\mathbf{K},n} \chi_{\mathbf{K},n}(\mathbf{r}_S),
\end{equation}
is given by
\begin{equation}
I=-2\sum_{n,m}c^\ast_{\mathbf{K},n}\widetilde{\Sigma}_{\mathbf{K},nm}
c_{\mathbf{K},m}.
\end{equation}

At each $\mathbf{K}$ and energy $E$ there are Bloch states travelling 
towards and away from the interface, as well as solutions decaying 
exponentially into the bulk \cite{heine}. 
These Bloch states and evanescent states 
can be considered as open and closed channels respectively, and
channels defined in this way have been used in transport studies
across interfaces \cite{wort,schep}.  In line with our 
convention, let us take the open Bloch channels to be the states 
propagating away from the interface, into the bulk; the states 
travelling in the opposite direction simply involve the perpendicular 
component of the wave-vector changing sign. We shall denote these 
Bloch states and the exponentially decaying states by 
$\phi_{\mathbf{K},i}(\mathbf{r})$, with $i$ labelling the states at
fixed $\mathbf{K}$ and $E$.

These Bloch states, evaluated over the interface, also diagonalize
$\widetilde{\Sigma}_{\mathbf{K}}$ (hence
$\Im\mathrm{m}\Sigma$), but we shall see that they are \emph{not} the
same as our channel functions $\psi_{\mathbf{K},i}$. As before, 
the flux associated with $\phi_{\mathbf{K},i}$ is given by
\begin{equation}
I_i=-2\int_S d^2r_S\int_S d^2r'_S \phi_{\mathbf{K},i}^*(\mathbf{r}_S)
\widetilde{\Sigma}_{\mathbf{K}}(\mathbf{r}_S,\mathbf{r}_S')
\phi_{\mathbf{K},i}
(\mathbf{r}_S'). \label{bloch1}
\end{equation}
There is also a well-known result \cite{wort}, 
which follows from Green's theorem
and the Bloch property of the wave-functions, that for $i\ne j$
\begin{equation}
\int_S
d^2r_S
\left[
\phi_{\mathbf{K},i}^*(\mathbf{r}_S)
\frac{\partial \phi_{\mathbf{K},j}}{\partial n_S}
(\mathbf{r}_S)-\phi_{\mathbf{K},j}(\mathbf{r}_S)
\frac{\partial \phi_{\mathbf{K},i}^*}{\partial n_S}(\mathbf{r}_S)
\right]=0,
\label{bloch2}
\end{equation}
hence
\begin{equation}
\int_S d^2r_S\int_S d^2r'_S \phi_{\mathbf{K},i}^*(\mathbf{r}_S)
\widetilde{\Sigma}_{\mathbf{K}}(\mathbf{r}_S,\mathbf{r}_S')
\phi_{\mathbf{K},j}
(\mathbf{r}_S')=0,~~~i\ne j.
\label{bloch3}
\end{equation}
Unfortunately, equations (\ref{bloch1}) and (\ref{bloch3}) do not 
imply that the $\phi_{\mathbf{K},i}$'s are eigenfunctions of 
$\widetilde{\Sigma}_{\mathbf{K}}$ or
$\Im\mathrm{m}\Sigma$. Although these functions are orthogonal over the
three-dimensional unit cell, it is known that the 
$\phi_{\mathbf{K},i}$'s are not orthogonal when integrated over 
the surface,
\begin{equation}
\int_S d^2r_S\phi_{\mathbf{K},i}^*(\mathbf{r}_S)\phi_{\mathbf{K},j}
(\mathbf{r}_S)\ne 0,~~~i\ne j.
\end{equation}
However, the $\phi_{\mathbf{K},i}$'s are linearly independent -- this
makes it possible to use them in surface and interface matching
procedures. In order for the function space spanned by the open 
$\phi_{\mathbf{K},i}(\mathbf{r}_S)$ and 
$\psi_{\mathbf{K},i}(\mathbf{r}_S)$ to be the same, there must be 
the same number of Bloch functions as our channel functions carrying 
finite flux.

The transformation matrix, which relates the flux-carrying Bloch
states with the open channel functions, has the very useful property 
of being unitary. Let us normalise the open 
$\phi_{\mathbf{K},i}(\mathbf{r}_S)$ and
$\psi_{\mathbf{K},i}(\mathbf{r}_S)$ to carry unit flux per
two-dimensional unit cell, away from the interface, so that
\begin{equation}
\int\! d^2r_S\int\! d^2r'_S \phi_{\mathbf{K},i}^*
\widetilde{\Sigma}_{\mathbf{K}}\phi_{\mathbf{K},j}=
-\frac{\delta_{ij}}{2},~
\int\! d^2r_S\int\! d^2r'_S \psi_{\mathbf{K},i}^*
\widetilde{\Sigma}_{\mathbf{K}}\psi_{\mathbf{K},j}=
-\frac{\delta_{ij}}{2}.
\label{ortho}
\end{equation}
We now expand the flux-carrying Bloch functions 
$\phi_{\mathbf{K},i}$, $\phi_{\mathbf{K},j}$
in terms of the open and closed channel functions (the closed channel 
functions can be normalised to unity using (\ref{norm})), 
\begin{equation}
\phi_{\mathbf{K},i}(\mathbf{r}_S)=\sum_{\mbox{\scriptsize open + closed}~m}
a_{im}\psi_{\mathbf{K},m}(\mathbf{r}_S),~~~
\phi_{\mathbf{K},j}(\mathbf{r}_S)=\sum_{\mbox{\scriptsize open + closed}~n}
a_{jn}\psi_{\mathbf{K},n}(\mathbf{r}_S). \label{exp1}
\end{equation}
Then from (\ref{ortho}) we obtain
\begin{eqnarray}
&&\int\! d^2r_S\int\!d^2r'_S\phi_{\mathbf{K},i}^*
\widetilde{\Sigma}_{\mathbf{K}}\phi_{\mathbf{K},j}=-\frac
{\delta_{ij}}{2}\nonumber \\ 
&&=\sum_{\mbox{\scriptsize open + closed}~m,n}
a_{im}^*a_{jn}\int\! d^2r_S\int\! d^2r'_S \psi_{\mathbf{K},m}^*
\widetilde{\Sigma}_{\mathbf{K}}\psi_{\mathbf{K},n}.\label{exp2}
\end{eqnarray}
But the integral on the right of (\ref{exp2}) vanishes when $m$ or $n$
is a closed channel, and when $m$, $n$ are both open the integral is
given by the second equation in (\ref{ortho}). Substituting, this
gives
\begin{equation}
\sum_{\mbox{\scriptsize open}~m}a_{im}^*a_{jm}=\delta_{ij},\label{exp3}
\end{equation}
showing that the matrix $a_{im}$ in the space of open functions is
unitary. 

As an example, we consider the case of embedding onto semi-infinite
Au(001), taking $\mathbf{K}=0$, zero wave-vector parallel to the
interface. The band-structure as a function of $k_\perp$, the
perpendicular component of the wave-vector, is shown in figure 2a,
corresponding to the $\Gamma$X direction in the bulk Brillouin
zone. For comparison, the figure also shows the eigenvalues of
\begin{figure}
\begin{center}
\epsfig{width=12cm,file=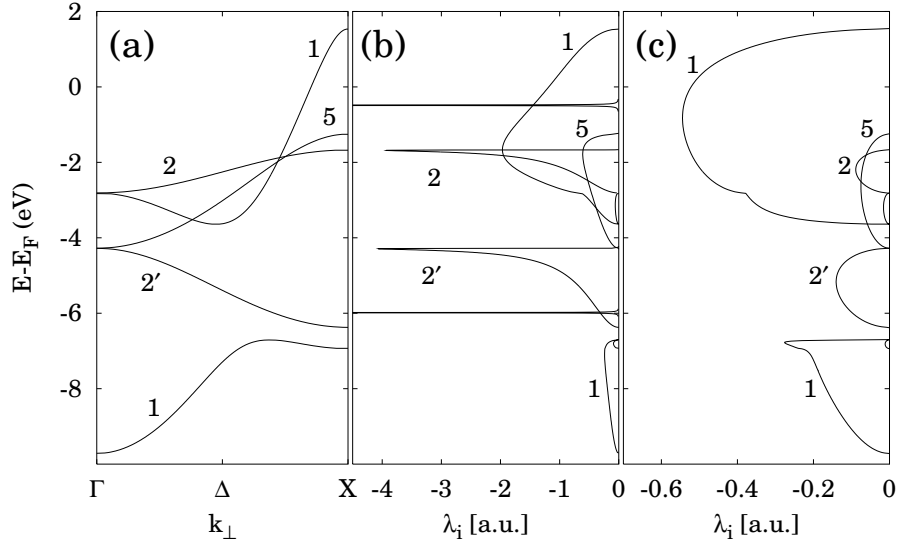}
\end{center}
\caption{Au(001) interface at $\mathbf{K}=0$: (a) band-structure in the
$\Gamma$X direction; (b) eigenvalues of
$\widetilde{\Sigma}_{\mathbf{K}=0}$ 
on an embedding plane midway between atomic planes. The calculation uses a 
finite imaginary energy of 0.1 meV; (c) eigenvalues of
$\widetilde{\Sigma}_{\mathbf{K}=0}$
on an embedding plane not midway between atomic planes.
The numbers label the symmetry of the corresponding eigenfunctions.}
\end{figure}
$\widetilde{\Sigma}_{\mathbf{K}}$
for $\mathbf{K}=0$ as a function 
of energy, in figure 2b where the embedding surface $S$ is 
midway between atomic planes, and in figure 2c 
where the embedding potential has been transferred to a plane
that does not intersect muffin-tin regions of the Au potential
\cite{crampin92,ishida}.
We see that the number of bands is indeed the same as
the number of non-zero eigenvalues, except at isolated energies where
there are $\delta$-function peaks. The $\delta$-function peaks will be
discussed later, but other features in the eigenvalue spectrum can be 
understood in terms of a one-dimensional nearly-free electron model. 
In this model the embedding potential near a band edge $E_0$ behaves 
like \cite{simon} 
\begin{equation}
\Im\mathrm{m}\Sigma \propto -\sqrt{|E-E_0|} \label{band1}
\end{equation}
if the amplitude of the wave-function at $E_0$ is finite on the 
interface. On the other hand, if the wave-function at $E_0$ is zero on
the interface, the embedding potential blows up at the band edge like
\begin{equation}
\Im\mathrm{m}\Sigma \propto -\frac{1}{\sqrt{|E-E_0|}}. \label{band2}
\end{equation}  
Even though this one-dimensional model cannot be directly applied to
the complicated bands of an fcc transition metal, our eigenvalues
behave like (\ref{band1}) or (\ref{band2}) at band extrema, exhibiting
the square root singularity on the embedding surface midway between
atomic planes (figure 2b) where symmetry permits the band edge 
wavefunction to vanish.

The interplay between the eigenvalues and the band-structure can be
quite complicated. The tiny semi-circle in the eigenvalue
figures, centred on $-6.8$ eV, varies in energy between the lowest 
X$_1$ state and the maximum of the first $\Delta_1$ band. For each 
energy in this range there are two non-zero eigenvalues, corresponding
to the two $\Delta_1$ Bloch states in the band structure. Similarly, 
there is a small eigenvalue semi-ellipse with energies
between the minimum of the second $\Delta_1$ band at $E=-3.7$ eV and
its maximum at $\Gamma_{12}$. It overlaps in energy with an outer
eigenvalue starting and finishing at $\lambda=0$, in the range of 
energies for which there are again two $\Delta_1$  Bloch states. 
This topology shows that when there is more than one 
$\phi_{\mathbf{K},i}$ at a particular energy with the same symmetry, 
there is not a one-to-one relationship with the $\psi_{\mathbf{K},i}$ 
channel functions. In the cases we have just described, the two 
$\Delta_1$ Bloch states are associated with one continuous band. 
This cannot have a one-to-one relationship with the two eigenvalues 
having the topology shown in figure 2. Rather, each Bloch state is a 
linear combination of all the channel functions at that energy, as 
given by (\ref{exp1}). Of course, in the case of a symmetry point 
like $\mathbf{K}=0$, the different bands at the same energy frequently
have different symmetry, and then we \emph{can} identify each band with 
one eigenvalue of $\widetilde{\Sigma}_{\mathbf{K}}$.

The $\delta$-function peaks are associated with discrete energies in a
symmetry gap, at $E=-6.0$ eV ($\Delta_1$ symmetry) and $-0.5$ eV
($\Delta_5$ symmetry). At these energies, the embedding potential has 
a singularity and its inverse $\Sigma^{-1}$ has a zero eigenvalue, 
which from (\ref{embpot1}) means that there is a normal derivative 
satisfying
\begin{equation}
\int_S d^2r'_S \Sigma^{-1}(\mathbf{r}_S,\mathbf{r}'_S)
\frac{\partial\psi(\mathbf{r}'_S)}{\partial n_S}=0. \label{delta}
\end{equation}
In other words, there is a wave-function in the system, with an
exponentially decaying envelope, with zero amplitude over the 
interface, and consequently infinite logarithmic derivative. This
gives the $\delta$-function peak in the eigenvalue of
$\widetilde{\Sigma}_{\mathbf{K}}$. This singularity only appears at an isolated
energy in the limit of $\widetilde{\Sigma}_{\mathbf{K}}$ being
evaluated at a real energy with an infinitesimal positive imaginary
part, and as the associated eigenvector $\psi_i(\mathbf{r}_S)=0$ there
is no flux associated with it.

\section{Channel functions and scattering}
We now use our channel states to provide a simple proof of 
(\ref{trans1}), starting off by deriving a general scattering result
containing $\Im\mathrm{m}\Sigma$. Let us consider the geometry shown
in figure 1 -- we wish to find the full wave-function 
$\chi(\mathbf{r})$ in region I, given an incoming incident wave 
$\psi(\mathbf{r})$ in the substrate region II. This can be done in
several ways starting from Green's theorem \cite{wort,dix,jeroen}, and
it follows straightforwardly that $\chi$ in region I is given by 
\begin{equation}
\chi(\mathbf{r})=\frac{1}{2}\int_S d^2r_S\left[G(\mathbf{r},
\mathbf{r}_S)\frac{\partial\psi}{\partial n_S}(\mathbf{r}_S)-
\frac{\partial G}{\partial n_S}(\mathbf{r},\mathbf{r}_S)
\psi(\mathbf{r}_S)\right],   \label{pert1}
\end{equation}
where $G$ is the Green function for the combined system of I and II;
the surface normal in (\ref{pert1}) is taken to be outwards from 
region I. Now because $G$ in (\ref{pert1}) is the outgoing Green 
function, we can use (\ref{embpot1}) to substitute for 
$\partial G/\partial n_S$ as follows
\begin{equation}
\frac{\partial G}{\partial n_S}(\mathbf{r},\mathbf{r}_S)=
-2\int_S d^2r_S' G(\mathbf{r},\mathbf{r}_S')\Sigma(\mathbf{r}_S',
\mathbf{r}_S). \label{pert2}
\end{equation}
The incident wave-function, on the other hand, satisfies
\begin{equation}
\frac{\partial\psi}{\partial n_S}(\mathbf{r}_S)=
-2\int_S
d^2r_S'\Sigma^*(\mathbf{r}_S,\mathbf{r}_S')\psi(\mathbf{r}_S'),
\label{pert3}
\end{equation}
the complex conjugation arising because this is an incoming wave.
Substituting (\ref{pert2}) and (\ref{pert3}) into (\ref{pert1}) gives
the pleasing result
\begin{equation}
\chi(\mathbf{r})=2i\int_S d^2r_S\int_S d^2r'_S G(\mathbf{r},
\mathbf{r}_S)\Im\mathrm{m}\Sigma(\mathbf{r}_S,\mathbf{r}_S')
\psi(\mathbf{r}_S'),~~~\mathbf{r}\mbox{ in I}.
\label{pert4}
\end{equation}

At this stage we must be clear about the incident 
wave $\psi$ in (\ref{pert4}). This can be any wave in region II in the
energy continuum (otherwise $\Im\mathrm{m}\Sigma$ in (\ref{pert4})
vanishes), satisfying incoming boundary conditions. If region II is a 
semi-infinite bulk crystal, for example, $\psi$ is an arbitrary
combination of the Bloch states approaching $S$, at fixed
energy, together with exponentially decaying states. The incident wave
is defined entirely by its value over $S$, $\psi(\mathbf{r}_S)$, which
can take an arbitrary form, provided that it has some projection onto
the flux-carrying states.

We now apply (\ref{pert4}) to a nanostructure connected to contacts 
over interfaces $S_l$ and $S_r$ (figure 1). In dividing the 
system in this way, the nanostructure region also contains the atoms 
of the contacts which are perturbed by the nanostructure. Region I 
consists of the whole structure to the right of $S_l$, that is, the 
nanostructure plus the right-hand contact and lead. If
$\psi(\mathbf{r})$ is the wave in the left-hand lead incident on
the nanostructure, the full wave-function everywhere to the right is 
given by (\ref{pert4}), with the double integral evaluated over
$S_l$. We take the incident wave to correspond to the open channel
function $\psi^l_i$ defined over the left-hand interface, normalised 
to unit flux over this interface. Then the wave-function over the 
right-hand interface is given by
\begin{equation}
\chi_i(\mathbf{r}_r)=2i\lambda^l_i\int_{S_l}d^2r_l G(\mathbf{r}_r,
\mathbf{r}_l)\psi^l_i(\mathbf{r}_l), \label{trans2}
\end{equation}
where $\lambda^l_i$ is the non-zero channel eigenvalue
(\ref{eigen}). We now expand $\chi_i(\mathbf{r}_r)$ in terms of the 
complete set of channel functions $\psi^r_j(\mathbf{r}_r)$ for the 
right-hand interface,
\begin{equation}
\chi_i(\mathbf{r}_r)=\sum_j t_{ij}\psi^r_j(\mathbf{r}_r), \label{trans3}
\end{equation}
and normalising the right-hand open channel functions to unit flux we
find, for open channels,
\begin{equation}
t_{ij}=4i\lambda^l_i|\lambda^r_j|
\int_{S_l}d^2 r_l\int_{S_r}d^2 r_r\psi^r_j
(\mathbf{r}_r)G(\mathbf{r}_r,\mathbf{r}_l)\psi^l_i(\mathbf{r}_l).
\label{trans4}
\end{equation}
The transmission coefficient $T_{ij}$ is the transmitted flux in the 
$j$th channel corresponding to unit incident flux in the $i$th
channel, and with our normalisation this is given by
\begin{equation}
T_{ij}=|t_{ij}|^2=16(\lambda^l_i\lambda^r_j)^2\left|
\int_{S_l}d^2 r_l\int_{S_r}d^2r_r\psi^r_j(\mathbf{r}_r)
G(\mathbf{r}_r,\mathbf{r}_l)\psi^l_i(\mathbf{r}_l)\right|^2.  
\label{trans5}
\end{equation}
This expression for the transmission coefficient is very simple, with
the clear interpretation of $G$ as a propagator from one channel to
the other. 

Let us now sum $T_{ij}$ over all the open exit channels, to give the 
total transmitted flux for unit incident flux in channel $i$. Changing
and simplifying the notation for spatial coordinates, this is given by
\begin{eqnarray}
\sum_{\mbox{\scriptsize open}~j}T_{ij}&=&16(\lambda_i^l)^2\int d1_r\int d2_l
\int d3_r\int d4_l G(2_l,1_r)\nonumber\\
&&\times\left(\sum_{\mbox{\scriptsize open}~j}
(\lambda_j^r)^2\psi_j^r(1_r)\psi_j^r(3_r)\right)
G^*(3_r,4_l)\psi_i^l(4_l)\psi^l_i(2_l).\label{trans6}
\end{eqnarray}
Now we have seen that $\Im\mathrm{m}\Sigma$ can be expanded in terms 
of the open channel eigenfunctions alone, and with unit flux
normalisation (\ref{eigsum2}) becomes
\begin{equation}
\Im\mathrm{m}\Sigma(\mathbf{r}_S,\mathbf{r}_S')=-2\sum_{\mbox{\scriptsize
open}~i}\lambda_i^2\psi_i(\mathbf{r}_S)\psi_i(\mathbf{r}_S').
\label{trans7}
\end{equation} 
Hence (\ref{trans6}) simplifies to
\begin{eqnarray}
\sum_{\mbox{\scriptsize open}~j}T_{ij}&=&-8(\lambda_i^l)^2\int d1_r\int d2_l
\int d3_r\int d4_l G(2_l,1_r)\Im\mathrm{m}\Sigma_r(1_r,3_r)\nonumber\\
&&\times G^*(3_r,4_l)\psi_i^l(4_l)\psi^l_i(2_l).\label{trans8}
\end{eqnarray}
Now we sum over all incident channels, to give
\begin{eqnarray}
\sum_{\mbox{\scriptsize open}~i,~j}T_{ij}&=&-8\int d1_r\int d2_l
\int d3_r\int d4_l G(2_l,1_r)\Im\mathrm{m}\Sigma_r(1_r,3_r)\nonumber\\
&&\times G^*(3_r,4_l)\left(\sum_{\mbox{\scriptsize open}~j}
(\lambda_i^l)^2\psi_i^l(4_l)\psi^l_i(2_l)\right),
\label{trans9}
\end{eqnarray}
and making use of (\ref{trans7}) once again, we obtain the result
\begin{eqnarray}
\sum_{\mbox{\scriptsize open}~i,~j}T_{ij}&=&4\int d1_r\int d2_l
\int d3_r\int d4_l G(2_l,1_r)\Im\mathrm{m}\Sigma_r(1_r,3_r)\nonumber\\
&&\times G^*(3_r,4_l)\Im\mathrm{m}\Sigma_l(4_l,2_l).
\label{trans10}
\end{eqnarray}
This gives the total transmission for unit flux in each of the
incident channels.

This result for total transmission is only useful if there is the same
flux in each open incident channel. To understand this let us consider
the case of semi-infinite interfaces, with unit flux in
each of the incident Bloch functions at fixed $\mathbf{K}$. Then the 
flux in the $m$'th channel function due to the flux in the $i$'th Bloch 
function is given in terms of the coefficients $a_{im}$ (\ref{exp1}) 
by $a^*_{im}a_{im}$. So the total flux in the $m$'th channel due to all
the open Bloch channels, each carrying unit flux, is given by $\sum_i 
a^*_{im}a_{im}$. But because $a_{im}$ is a unitary matrix, we have 
$\sum_i a^*_{im}a_{im}=1$, and unit flux in each Bloch function
implies unit flux in each channel function. This is the result we 
require: with energy normalisation, the Bloch functions -- hence the 
channel functions -- all carry the same flux. We can then write the 
total transmission between left and right as
\begin{eqnarray}
T_{lr}&=&4\int d1_r\int d2_l
\int d3_r\int d4_l G(2_l,1_r)\Im\mathrm{m}\Sigma_r(1_r,3_r)
G^*(3_r,4_l)\Im\mathrm{m}\Sigma_l(4_l,2_l)\nonumber \\
&=&4\textrm{Tr}[G_{lr}\Im\mathrm{m}\Sigma_rG^*_{rl}\Im\mathrm{m}\Sigma_l],
\label{trans11}
\end{eqnarray}
recovering (\ref{trans1}). This result is manifestly independent of
choice of channel functions.

To summarise, we have shown how the eigenfunctions of the embedding potential, 
a generalised logarithmic derivative, may be used to define 
conduction channels across a surface. These new channel 
functions are orthogonal, which could instil distinct advantages in 
applications compared to Bloch states as conduction channels. Using the new
channel functions we have provided a simple derivation of a well-known
result for the total transmission through a conductor system.

\section*{Acknowledgements}
We wish to thank Owen Davies, Ian Merrick, Sir John Pendry, and 
Joshua Zak for their help and suggestions in this work.

\end{document}